\documentclass[11pt]{article}
\pdfoutput=1
\newcommand{\Comment}[1]{{}}
\usepackage{amsfonts,amsthm,amsmath,amssymb,slashed}
\usepackage[textwidth = 430 pt, textheight = 630 pt]{geometry}
\usepackage{color}
\usepackage{booktabs}
\usepackage{subcaption}
\usepackage[export]{adjustbox}

\Comment{\usepackage{color}
\definecolor{MyDarkBlue}{rgb}{0.15,0.15,0.45}
\usepackage[linktocpage=true]{hyperref}
\hypersetup{
colorlinks=true,
citecolor=MyDarkBlue,
linkcolor=MyDarkBlue,
urlcolor=MyDarkBlue,
pdfauthor={Horatiu Nastase and Caio Luiz Tiedt},
pdftitle={Holographic transport with topological term and entropy function},
pdfsubject={hep-th}
}
\usepackage[utf8]{inputenc}
\usepackage[numbers,sort&compress]{natbib}
\usepackage{hypernat}}
\usepackage{graphicx}
\usepackage{cite}
\usepackage[colorlinks=true]{hyperref}
\hypersetup{
  allcolors = blue,
}

\newcommand\ignore[1]{}
\def\one{{\,\hbox{1\kern-.8mm l}}}

\def\a{\alpha}\def\b{\beta}

\def\d{\partial}

\newcommand{\Cset}{{\,\,{{{^{_{\pmb{\mid}}}}\kern-.45em{\mathrm C}}}}}

\newcommand{\be}{\begin{equation}}
\newcommand{\bea}{\begin{eqnarray}}

\newcommand{\ee}{\end{equation}}
\newcommand{\eea}{\end{eqnarray}}

\begin{document}

\renewcommand{\thefootnote}{\fnsymbol{footnote}}

\makeatletter
\@addtoreset{equation}{section}
\makeatother
\renewcommand{\theequation}{\thesection.\arabic{equation}}

\rightline{}
\rightline{}




\begin{center}
{\LARGE \bf{\sc Holographic transport with topological term and entropy function}}
\end{center}
 \vspace{1truecm}
\thispagestyle{empty} \centerline{
{\large \bf {\sc Horatiu Nastase${}^{a}$}}\footnote{E-mail address: \Comment{\href{mailto:horatiu.nastase@unesp.br}}{\tt horatiu.nastase@unesp.br}}
{\bf{\sc and}}      
{\large \bf {\sc Caio Luiz Tiedt${}^{b}$}}\footnote{E-mail address: \Comment{\href{mailto:caio.tiedt@if.usp.br}}{\tt caio.tiedt@if.usp.br}}                                           }

\vspace{.5cm}


\centerline{{\it ${}^a$Instituto de F\'{i}sica Te\'{o}rica, UNESP-Universidade Estadual Paulista}}
\centerline{{\it R. Dr. Bento T. Ferraz 271, Bl. II, Sao Paulo 01140-070, SP, Brazil}}
\vspace{.3cm}
\centerline{{\it ${}^b$Instituto de F\'{i}sica, Universidade de S\~ao Paulo,}}
\centerline{{\it Rua do Mat\~ao, 1371, 05508-090 S\~ao Paulo, SP, Brazil}}

\vspace{1truecm}

\thispagestyle{empty}

\centerline{\sc Abstract}

\vspace{.4truecm}

\begin{center}
\begin{minipage}[c]{380pt}
{\noindent In this paper we consider the effect of a topological Maxwell term $W(\Phi)F_{\mu\nu}\tilde 
F^{\mu\nu}$ on holographic transport and thermodynamics in 2+1 dimensions, 
in the case with a dyonic black hole in the gravity dual. We find that for a constant $W$ the modifications
to the thermodynamics are easily quantified, and transport is affected only for $\sigma_{xy}$. 
If one consider also 
the attractor mechanism, and writing  the horizon transport in terms of charges, the transport
coefficients are affected explicitly. We also introduce the case of case of 
radially dependent $W(z)$, in which case, however, analytical calculations become very involved.
We also consider the implications of the two models for the S-duality of holographic transport coefficients.
}
\end{minipage}
\end{center}

\vspace{.5cm}

\setcounter{page}{0}
\setcounter{tocdepth}{2}

\newpage

\renewcommand{\thefootnote}{\arabic{footnote}}
\setcounter{footnote}{0}

\linespread{1.1}
\parskip 4pt



\section{Introduction}

The AdS/CFT correspondence \cite{Maldacena:1997re} (see \cite{Nastase:2015wjb,Ammon:2015wua} 
for a review) describes strongly coupled phenomena in field theories using perturbative string theory or, 
more commonly, its classical (super)gravity limit. Applications to condensed matter via AdS/CMT
(see \cite{Nastase:2017cxp} for a review), are 
(usually) phenomenological in nature: one defines a gravity theory which, via the holographic map that 
is still presumed to work, have expected properties for the dual strongly coupled field theory. 
One of the most standard applications is for electric and thermal transport in various materials at 
nonzero temperature, dual to properties in the background of a black hole 
(see \cite{Hartnoll:2009sz,Nastase:2017cxp} for a review). 

One set of methods was considered, more recently, in  \cite{Blake:2015ina,Donos:2015bxe,Donos:2014cya,Banks:2015wha,Donos:2014uba,Donos:2017mhp,Erdmenger:2016wyp}, where one considers 
an Einstein-Maxwell-dilaton ($g_{\mu\nu}, A_\mu, \phi$) system in the gravity dual, 
also coupled to axions ($\chi_1,\chi_2$), 
that allows for a more general analysis of transport, from the horizon of a black hole, based on the 
application of the membrane paradigm to AdS/CFT in the form started in \cite{Iqbal:2008}, relating 
the horizon to the boundary, where standard AdS/CFT quantities are obtained, via a radial evolution 
equation. In \cite{Alejo:2019hnb,Alejo:2019utd,Melnikov:2020ktj}, 
an extension of the analysis was considered, by adding
a topological $W(\phi)F_{\mu\nu}\tilde F^{\mu\nu}$ term and considering the resulting S-duality 
properties. One can also use the attractor mechanism and Sen's entropy function 
\cite{Sen:2005wa} to calculate conductivities from black hole horizons \cite{Erdmenger:2016wyp}, 
and this method was also used in \cite{Alejo:2019utd}.

But the original analysis of transport (for shear viscosity $\eta$ of a strongly coupled field theory 
fluid), defined in \cite{Son:2002sd,Policastro:2002se,Kovtun:2004de}, was used earlier for calculating 
the thermo-electric transport from a dyonic black hole in the gravity dual in 
\cite{Hartnoll:2007ai,Hartnoll:2007ih}, with a more general analysis in \cite{Davison:2016ngz}
(see also, for instance, more recently \cite{Sachdev:2019bjn}). 

In this paper we are interested in considering the effect of a $W(\phi)$ topological term on the 
calculation of thermodynamics and tranport in \cite{Hartnoll:2007ai,Hartnoll:2007ih}, as well as the 
calculations via the attractor mechanism. 
The presence of the $W(\phi)$ term is necessary in order to obtain complete S-duality transformations 
in the gravity dual, and they correspond to $Sl(2,\mathbb{Z})$ transformations on observables in the 
gauge theory, including particle-vortex duality \cite{Alejo:2019hnb,Alejo:2019utd,Melnikov:2020ktj}, both of which are of great 
theoretical importance in condensed matter systems. 
We will first consider the case of a constant $W(\phi)$ and then, 
in order to obtain more nontrivial results, we try to consider varying $W(\phi)$. However, it turns out that 
the simplest possible model is to consider directly a fixed $W(\phi(z))=W(z)$, with $z$ the radial direction, 
since otherwise solving the equations is very difficult. 

The paper is organized as follows. In section 2, we consider the topological term with constant 
coefficient $W$, calculating the thermodynamics, the holographic transport coefficients $\sigma_{ab}$ and 
$\a_{ab}$ from the Kubo formulas at the boundary, and then the attractor mechanism using Sen's 
entropy function, to write these in terms of charges and parameters at the horizon. The action of S-duality 
on the holographic model and the transport coefficients is also explained.
In section 3, we consider a ``toy model'' for a field dependent $W(\phi)$, where both $W$ and $Z$ 
(the coefficient of the Maxwell term) are instead explicit functions of the radial coordinate $z$. We 
find the solutions for fluctuations, though the holographic transport coefficients are too complicated to 
write, as are the results of the entropy function formalism. The effect of S-duality is explained, as is 
the introducing of anisotropy in the model. In section 4 we conclude, and the Appendices contain some 
long formulas.

\section{Thermoelectric conductivities from Kubo formulas, with topological term}

We want to extend the analysis of thermodynamics and transport from fluctuations found in 
\cite{Hartnoll:2007ai,Hartnoll:2007ih} to the presence of the topological term. As there, we will consider
a full solution for a dyonic black hole, but otherwise, like \cite{Alejo:2019utd,Melnikov:2020ktj},
where only the horizon was considered (still with nonzero magnetic field), we will add the topological term. 

Consider the action for gravity and a Maxwell field given by 
\begin{equation}\label{Action}
    I = -\frac{4}{2\kappa_4^2}\int d^4 x \sqrt{-g} \left(-\frac{1}{4}R + \frac{L^2}{4}F_{\mu\nu}F^{\mu\nu} - \frac{3}{2}\frac{1}{L^2} + W \frac{L^2}{4}F_{\mu\nu}\tilde{F}^{\mu\nu}\right)\;,
\end{equation}
where $F_{\mu\nu}$ is the field strength of $A_\mu$, $ F_{\mu\nu} = \partial_\mu A_\nu 
- \partial_\nu A_\mu$, and $\tilde F_{\mu\nu}$ is its dual, 
\be
\tilde{F}_{\mu\nu}=\frac{1}{2}  \sqrt{-g}\epsilon_{\mu\nu\rho\sigma}  F^{\rho\sigma}.
\ee

The normalization of the Newton's constant $\kappa_4$ relative to the radius  $L$ for the (negative) 
cosmological constant is given in terms of the rank $N$ of the dual field theory by 
\be
\frac{2L^2}{\kappa^2_4} = \frac{\sqrt{2}N^{3/2}}{6\pi}.
\ee

Note that with respect to the more general formulas in \cite{Alejo:2019utd,Melnikov:2020ktj} we have set
\be
\frac{Z}{g_4^2}=\frac{L^2}{4\pi G_N}\equiv \frac{c}{\pi}\;,\;\;\;  W(\phi)=W \frac{L^2}{16\pi G_N}\;,\label{conv}
\ee
in order to match with the analysis of the dyonic black hole in \cite{Hartnoll:2007ai,Hartnoll:2007ih}, and 
to simplify the analysis of the thermodynamics. 

The equations of motion for gravity and the Maxwell field are 
\begin{equation}\label{eomGravity}
    R_{\mu\nu}  = 2L^2F_{\mu\sigma}F_\nu^\sigma - \frac{L^2}{2}g_{\mu\nu}F_{\sigma\rho}F^{\sigma\rho} -\frac{3}{L^2}g_{\mu\nu},
\end{equation}    

\begin{equation}\label{EOMGAUGE}
    \nabla_\mu \left(F^{\mu\nu} + W\tilde{F}^{\mu\nu} \right) = 0.
\end{equation}

However, since 
\be
\nabla_\mu F^{\mu\nu} = \frac{1}{\sqrt{-g}} \partial_\mu (\sqrt{-g})F^{\mu\nu}\;,
\ee
the Maxwell equation becomes
\be
\frac{1}{\sqrt{-g}} \partial_\mu \sqrt{-g}\left(F^{\mu\nu} + W\tilde{F}^{\mu\nu} \right) = 0.
\ee

On the other hand, by its topological nature, the $W$ term doesn't affect the Einstein equation. 
Therefore we can use the same solution as in \cite{Hartnoll:2007ai,Hartnoll:2007ih}, 
\be
 \frac{1}{L^2}ds^2 = -\frac{\alpha^2}{z^2}f(z)dt^2+\frac{\alpha^2}{z^2}(dx^2+dy^2)+\frac{1}{z^2}\frac{dz^2}{f(z)}\;,
\ee
where $f(z)$ is written in terms of (properly normalized) magnetic charges $h$ and electric charges $q$
as
\be
 f(z) = 1+(h^2+q^2)z^4-(1+h^2+q^2)z^3.
\ee

Here the horizon of the black hole is at $z=1$, and the boundary of the asymtotic AdS space in 
which it is embedded is at $z\rightarrow 0$. Note that $\sqrt{-g}=L^4\a^3/z^4$.

Since $W$ is constant, the solution for the gauge field $A_\mu$ is also the same, 
\be
 A = h\alpha^2xdy+q\alpha (z-1)dt\;,\label{Asol}
\ee
where the gauge choice was such that $A_t\rightarrow 0$ at the horizon $z\rightarrow 1$, as is 
usual for AdS/CFT. Then we also have 
\be
F=h\a^2dx\wedge dy +q \a\, dz\wedge dt\;,\label{Fsol}
\ee
or, explicitly in components $(t,x,y,z)$, 
\be
F_{\mu\nu} = 
\begin{pmatrix}
0 & 0 & 0 & -q\alpha\\
0 & 0 & h\alpha^2 & 0\\
0 & -h\alpha^2 & 0 & 0\\
q\alpha & 0 & 0 & 0
\end{pmatrix}\;,\;\;\;
\tilde{F}^{\mu\nu} = \frac{z^4}{L^4\alpha^3}
\begin{pmatrix}
0 & 0 & 0 & h\alpha^2\\
0 & 0 & -q\alpha & 0\\
0 & q\alpha & 0 & 0\\
-h\alpha^2 & 0 & 0 & 0
\end{pmatrix}.
\ee

\subsection{Thermodynamics}

The temperature of the black hole is, as usual, 
\be
 T=\frac{\left.\alpha f^\prime (1/z)\right|_{z=1}}{4\pi} = \frac{\alpha (3-h^2-q^2)}{4\pi}.\label{temp}
\ee

To compute the thermodynamics, we note that we are in the grand canonical ensemble, since we have 
fixed charges, so $\Omega=TI_{\rm ren}$. 
However, the action needs to be renormalized by the addition of counterterms,
\be
I_{\rm ren}=I+I_{\rm ct}\;,
\ee
and we can use the standard counterterm action from the boundary of AdS, 
\be
 I_\text{ct} = -\frac{1}{\kappa^2_4}\int d^3x \sqrt{-\gamma}\theta -\frac{2}{\kappa^2_4}\frac{1}{L}\int d^3x \sqrt{-\gamma}.\label{Counterterms}
\ee

The unit normal to the boundary, satisfying $n^\mu n_\mu=1$, is 
\be
n^\mu = z\frac{\sqrt{f(z)}}{L}\times (0,0,0,1).
\ee

The boundary metric $\gamma_{MN}$ is the one induced by $g_{\mu\nu}$ in the 3 boundary 
directions, but since it blows up as $z\rightarrow 0$, it must be defined a bit away from the 
boundary, at $z=\epsilon$ with $\epsilon\rightarrow 0$, obtaining 
\be
 \gamma_{MN} = -\frac{\alpha^2}{\epsilon^2}f(\epsilon)dt^2+\frac{\alpha^2}{\epsilon^2}(dx^2+dy^2).
\ee

Equivalently, we then have $\gamma_{\mu\nu}=g_{\mu\nu}-n_\mu n_\nu$. 

The $\theta$ appearing in the counterterm action is the trace of the extrinsic curvature
\be
\theta_{MN} = -\frac{1}{2}(\nabla_\mu n_\nu + \nabla_\nu n_\mu) = \begin{pmatrix}
\frac{L\sqrt{f(\epsilon)}(\epsilon f^\prime (\epsilon)-2f(\epsilon))}{2\epsilon^2} & 0 & 0 \\
0 & -\frac{L\sqrt{f(\epsilon)}\alpha^2}{\epsilon^2} & 0 \\
0 & 0 & -\frac{L\sqrt{f(\epsilon)}\alpha^2}{\epsilon^2}
\end{pmatrix}\;,
\ee
with 
\be
 \theta = \gamma^{MN}\theta_{MN} = \frac{\epsilon f^\prime (\epsilon ) -6f(\epsilon ) }
 {2L\sqrt{f(\epsilon )}}.
\ee

Plugging into the counterterm action (\ref{Counterterms}), and doing the integral over time $t$ up 
to the inverse temperature $\b$, we get, in a power series in $\epsilon$, 
\be
 I_\text{ct} = \beta V \frac{\sqrt{2}N^{3/2}}{6\pi}\frac{\alpha^3}{2} \left( \frac{-1}{\epsilon^3} + \frac{1+h^2+q^2}{2} +\mathcal{O}(\epsilon ) \right)\;,
\ee
where $V=\int dxdy$.

Integrating $z$ in the action $I$ from $\epsilon$ (the regularized boundary) to 1 (the horizon), we 
obtain
\be
I = \beta V \frac{\sqrt{2}N^{3/2}}{6\pi}\frac{\alpha^3}{2}\left( \frac{1}{\epsilon} -1+h^2-q^2-2Whq +\mathcal{O}(\epsilon )\right)\;,
\ee
so that the renormalized action becomes
\be
I_{\rm ren} = \lim_{\epsilon\to 0} (I + I_{\rm ct}) = \beta V \frac{\sqrt{2}N^{3/2}}{6\pi}\frac{\alpha^3}{4}\left( -1-q^2+3h^2 -4Whq\right).
\ee

Multiplying with the temperature gives the grand canonical potential, 
\be
\Omega = \frac{\sqrt{2}N^{3/2}}{6\pi}\frac{\alpha^3V}{4}\left( -1-q^2+3h^2 -4Whq\right).
\ee

The magnetic field and the chemical potential are found from the $z\rightarrow 0$ limit of $A$ in 
(\ref{Asol}) and its field strength $F$ in (\ref{Fsol}), 
\be
B=\a^2h\;,\;\;\;\;\; \mu =-\a q.
\ee

In terms of these (its natural) parameters, the grand canonical potential is 
\be
\Omega = \frac{\sqrt{2}N^{3/2}}{6\pi}\frac{\alpha^3V}{4}\left( -1-\frac{\mu^2}{\alpha^2}+3\frac{B^2}{\alpha^4} +4W\frac{\mu B}{\alpha^3}\right).
\ee

While this potential depends on $W$, the entropy $S$ and energy $E$ derived from it must not, since 
they depend only on the geometry of the black hole. Indeed, for the entropy, we first find from the above 
$\Omega$ and the $T$ in (\ref{temp}) that
\be
\left.dT\right|_{B,\mu}=\frac{d\a}{4\pi}\left(3+3\frac{B^2}{\a^4}+\frac{\mu^2}{\a^2}\right)\;,
\;\;\;
\left.d\Omega\right|_{B,\mu}= \frac{\sqrt{2}N^{3/2}}{6\pi}\frac{3\alpha^2d\a V}{4}\left(-3-
\frac{\mu^2}{\a^2}-3\frac{B^2}{\a^4}\right)\;,
\ee
so that 
\be
S=-\left.\frac{\d\Omega}{\d T}\right|_{B,\mu}=\frac{\sqrt{2}N^{3/2}}{6\pi}\alpha^2V=c\alpha^2 V.
\ee
This indeed matches the entropy calculated from the area of the horizon, with our conventions in (\ref{conv}).

Then the energy is 
\be
E=\Omega+TS+\mu N=\frac{\sqrt{2}N^{3/2}}{6\pi}\frac{\alpha^3V}{2}\left( 1+q^2+h^2 \right)\;,
\ee
where $N=Q$ (the electric charge) is calculated below: 

Indeed, from (\ref{temp}), we find that when $T, B$ are fixed, we get $\mu=\mu(\a)$, as
\be
d\a\left(3+\frac{\mu^2}{\a^2}+3\frac{B^2}{\a^4}\right)=2\mu \frac{d\mu}{\a}\;,
\ee
and then 
\be
\left.d\Omega\right|_{T,B}
= \frac{\sqrt{2}N^{3/2}}{6\pi}\frac{V}{4}\left\{d\a\left[-3\a^2-\mu^2-3\frac{B^2}{\a^2}\right]
+d\mu\left[-2\a\mu+4BW\right]\right\}\;,
\ee
so that 
\be
Q=-\left.\frac{\d \Omega}{\d \mu}\right|_{T,B}= \frac{\sqrt{2}N^{3/2}}{6\pi}\frac{V}{4}(4\a \mu
-4BW).
\ee

Again from (\ref{temp}), when $T,\mu$ are fixed, we get $\a=\a(B)$,
\be
d\a\left(3+\frac{\mu^2}{\a^2}+3\frac{B^2}{\a^4}\right)=2B\frac{dB}{\a^3}\;,
\ee
and then 
\be
\left. d\Omega\right|_{T,\mu}=\frac{\sqrt{2}N^{3/2}}{6\pi}\frac{V}{4}\left\{d\a\left[-3\a^2-\mu^2
-3\frac{B^2}{\a^2}\right]+6\frac{BdB}{\a}+4W\mu dB\right\}\;,
\ee
so that
\be
M=\left.\frac{d\Omega}{dB}\right|_{T,\mu}=\frac{\sqrt{2}N^{3/2}}{6\pi}\frac{V}{4}\left(\frac{4B}{\a}
+4W\mu\right).
\ee

So the charge and magnetization do get a $W$ term, while the entropy and energy don't.

\subsection{Magnetization and fluctuations}

We want to use the Kubo formulas to calculate transport coefficients, and for that we need to calculate 
the holographic Green's functions, from fluctuations in the AdS dyonic black hole background. 

We follow the prescription from \cite{Hartnoll:2007ai,Hartnoll:2007ih}. But as described there, 
and found in \cite{Cooper1997}, in the case 
of nonzero magnetic field, we must subtract the magnetization currents from the standard thermo-electric
ones, 
\begin{equation}
    J^{\text{mag}}_i = \epsilon_{ij}\partial_j M,
\end{equation}
\begin{equation}
    T^{\text{mag}}_{ti} = \epsilon_{ij}\partial_j M^E,
\end{equation}
where $M$ and $M^E$ are respectively the magnetization and the energy magnetization densities. 

We must consider fluctuations on top of the background, that will give the holographic Green's functions, 
compatible with the boundary condition in terms of a magnetic field $B$ and a ``energy magnetic 
field $B^E$'' (sometimes called $B_1$, that sources $g_{ty}$ instead of $A_y$). 
Since $\delta A_\mu^0$ couples to a source current $J_\mu$ nd $\delta g_{\mu\nu}^0$ 
couples to a source energy-momentum tensor $T_{\mu\nu}$, we must have the boundary condition 
\bea
\lim_{z\to 0}\delta A_y &=& xB\;,\cr
\lim_{z\to 0}\delta G_y &=& xB^E\;,
\eea
where $G_y = g_{ty}z^2/\alpha$.

The ansatz compatible with these boundary conditions is \cite{Hartnoll:2007ih}
\bea
\delta A_y &=& x(B-qB^Ez)\;,\cr
 \delta G_y &=& xf(z)B^E.
\eea

From the equations of motion, it follows also that we need (only) $\delta A_t$ to compensate the above. 
We then construct the on-shell quadratic action for the fluctuations. 

To find it, we introduce a parameter $\epsilon$ to keep track of fluctuations to linear order (linear 
response), 
\bea
g_{\mu\nu} &=& g^{\text{background}}_{\mu\nu} + \epsilon \delta g_{\mu\nu}\;,\cr
A_\mu &=& A^{\text{background}}_\mu + \epsilon \delta A_\mu\;,
\eea
and solve the equations of motion in an expansion in $\epsilon$. The zeroth order term is just the 
equation of motion for the background, so vanishes, while the Einstein equation 
\be
 0 = 2L^2F_{\mu\sigma}F_\nu^\sigma - \frac{L^2}{2}g_{\mu\nu}F_{\sigma\rho}F^{\sigma\rho} -\frac{3}{L^2}g_{\mu\nu} - R_{\mu\nu}
\ee
at order $\epsilon$, for $(\mu\nu)=(xx)$, for example, gives
\be
 2z^2(Bh-B^Eqhz+q\alpha \delta A_t^\prime (z))\epsilon + \mathcal{O}(\epsilon^2)\;,
\ee
where prime denotes differentiation with respect to $z$. Then the linear order for the equations of 
motion fixes the $\delta A_t$ fluctuation, 
\be
\delta A_t^\prime (z) = -\frac{Bh-B^Eqhz}{q\alpha}.\label{Atprime}
\ee

It then turns out that this solves the Einstein equations at linear order in $\epsilon$ for all $(\mu\nu)$. 

On the other hand, the Maxwell equation in the absence of $W$ is expanded in $\epsilon$ 
as 
\be
\nabla_\mu F^{\mu t}  = \frac{z^4}{\alpha^3}(B^Eh-\alpha \delta A_t^{\prime\prime} (z))\epsilon + \mathcal{O}(\epsilon^2)\;,
\ee
giving 
\be
\delta A_t^{\prime\prime} (z) = \frac{B^Eh}{\alpha}.\label{Atdoublep}
\ee

The solution of (\ref{Atprime}) and (\ref{Atdoublep}) is 
\be
\delta A_t (z) = \frac{hB^E}{2\alpha}(z^2-1) - \frac{hB}{q\alpha}(z-1)\;,
\ee
where the integration constants were fixed such that $\delta A_t\rightarrow 0$ at the horizon $z=1$. 

In the presence of the topological term with $W$, since $W$ is a constant, the extra term vanishes 
since 
\be
 \nabla_\mu \tilde{F}^{\mu \nu} = 0\;,
\ee
so for a modification of the action for fluctuations we would need to consider the case of a $W$ that 
depends on $z$, which will be done later.

\subsection{Transport coefficients from quadratic action at boundary}

Here we find  the thermo-electric coefficients from the quadratic boundary action.

The (retarded) Green's function is found from the on-shell quadratic action \cite{Policastro:2002se}. 
Consider as an 
example a scalar field $\phi$ with a kinetic function $A(z)$ at the 2+1 dimensional boundary, 
\be
 S_{\rm cl} = \frac{1}{2}\int dz d^2x A(z) (\partial_\mu \phi )^2\;,
\ee
where we have solved the linearized field equations in terms of the boundary value $\phi_0$ in frequency
space $\omega$,
\be
 \phi(z) = f_\omega(z) \phi_0\;,
\ee
such that $f_\omega(z)=1$, and impose incoming wave boundary conditions at the horizon. 

Then the retarded Green's function is 
\be
G^R = \lim_{z\rightarrow 0} A(z)f_ {-\omega}(z)\partial_zf_\omega.
\ee

A more precise recipe is to go to momentum space $\vec{k}$ in all boundary directions, and 
write the on-shell action as 
\be
S_{\rm on-shell}=-\left.\int \frac{d^3k}{(2\pi)^3}\phi_0(-\vec{k})\frac{1}{2}{\cal F} (\vec{k},z)
\phi_0(\vec{k})\right|_{z=0}^{z=1}\;,
\ee
which identifies ${\cal F}$ as $G^R$. 

Either way, in our case, we consider frequency-space fluctuations
\be
\delta A_i=A_i(z) e^{i \omega t}\;,\;\;\;\;
\delta G_i =G_i(z) e^{i\omega t}\;,
\ee
for $i=x,y$ and $\a G_i =g_{ti}z^2$, and compute the linearized Maxwell equations, giving 
\bea
f(f A_x^\prime)^\prime + \bar{\omega}A_x + i\bar{\omega}hG_y + qfG^\prime_x &=& 0\cr
f(f A_y^\prime)^\prime + \bar{\omega}A_y + i\bar{\omega}hG_x + qfG^\prime_y &=& 0\;,\label{flucteq}
\eea
where $\bar \omega=\omega/\a$.

The Einstein equations give the same equations. 

Since near the horizon $f(z)\sim 1-z$, and near the horizon we obtain the solutions $A_i(z)\sim (1-z)^\nu$
and $G_i(z)\sim (1-z)^{\nu +1}$, we can write the solutions that are incoming at the horizon as 
\bea
A_i(z) &= & f(z)^\nu a_i(z)\cr
G_i(z) &= & f(z)^{(1+\nu)}g_i(z)\;,
\eea
where $\nu=i\bar{\omega}/(h^2+q^2-3)$ and the functions $a_i(z)$ and $g_i(z)$
are required to be regular at the horizon $z=1$. 

Note that we also have solutions with $\nu \rightarrow -\nu$ at the horizon, as well as the constant 
solution 
\be
G_y =  \frac{i\bar{\omega}}{h}A_x\;,\;\;\;
G_x =  -\frac{i\bar{\omega}}{h}A_y.
\ee

For the {\em static } transport coefficients we are interested in the hydrodynamic limit $\bar\omega
\rightarrow 0$. For this, we expand the fluctuations at constant dyonic charges $h$ and $q$, 
\bea
a_i(z) &= & a^{(0)} + \bar{\omega}a_i^{(1)}(z) + ...,\cr
g_i(z) &= & g^{(0)} + \bar{\omega}g_i^{(1)}(z) + ...
\eea
and solve (\ref{flucteq}) to first order, obtaining 
\bea
G_x(z) &=& -\frac{i\bar{\omega}}{h}\delta_y + f(z)^{1+\nu}\left( G_x^0 + \frac{i\bar{\omega}}{h}
\delta_y - i\bar{\omega}G_x^0\int_0^z \frac{du}{\psi^2(u)}P_5(u) \right)\;,\cr
A_x(z) &=& \delta_x +f(z)^\nu  \left( A^0_x - \delta_x - \left(G^0_x +  \frac{i\bar{\omega}}{h}\right)qz 
+ i\bar{\omega}q G^0_x \int^z_0 \frac{du (z-u)}{\psi^2(u)} P_5(u) \right.\cr
    && \hspace{97pt} \left. - i\bar{\omega} \int_0^z\frac{du}{f(u)}(G^0_xQ_4(u)+G^0_yQ_3(u))  
    \right).\label{onshellfluct}
\eea

Here $\psi(z) = f(z)/z$, the functions $P_5(u)$, $Q_3(u)$ and $Q_4(u)$ are polynomials found in
Appendix 1 of \cite{Hartnoll:2007ai}, and $\delta_i$ are constants that depend 
on the boundary values of the fields,
\begin{equation}
    \delta_x = A^0_x + \frac{G^0_y h (h^2+q^2-3)- 3 G^0_x q (1+h^2+q^2)}{4(h^2+q^2)}\;,
\end{equation}
and similarly for $\delta_y$. Note that $G_x$ has a $\delta_y$ component, so a $A^0_y$ component, 
and similarly $G_y$ has a $A^0_x$ component (thus mixing of $x$ with $y$ components), while $A_x$ has 
only an $A_x^0$ component (no mixing).

For the quadratic action for fluctuations, 
we derive first the gauge field terms (with $A_x, A_y$), which come from 
the Maxwell and topological terms in the action, 
\begin{equation}
\begin{split}
    \frac{2}{\kappa_4^2}\int d^4 & x \sqrt{-g}  \left(\frac{L^2}{4}F_{\mu\nu}F^{\mu\nu} + W \frac{L^2}{4}F_{\mu\nu}\tilde{F}^{\mu\nu}\right)  \rightarrow  \\
    &  \frac{\alpha L^2}{\kappa_4^2}\int d^4 x  \left( \frac{\bar{\omega}}{f(z)}(A_x^2+A_y^2) + f(z) ((A_x^\prime)^2+(A_y^\prime)^2) + 2 i W\bar{\omega}(A_x^\prime A_y - A_x A_y^\prime) \right).
\end{split}    
\end{equation}

Integrating by parts and neglecting terms calculated at the horizon $(z=1)$, 
and focusing on the lowest order in $\omega$ in each of the two terms (Maxwell and topological)
separately, we obtain 
\begin{equation}
\begin{split}
    &\frac{2}{\kappa_4^2}\int d^4  x \sqrt{-g}  \left(\frac{L^2}{4}F_{\mu\nu}F^{\mu\nu} 
    + W \frac{L^2}{4}F_{\mu\nu}\tilde{F}^{\mu\nu}\right)  \rightarrow  \\
    &  \frac{\alpha L^2}{\kappa_4^2}\int d^3 x  \lim_{z\rightarrow 0} \left[f(z) (A_x^\prime A_x 
    + A_y^\prime A_y)+2i\bar\omega W\epsilon^{ab}A'_a A_b\right].  
\end{split}    
\end{equation}

Indeed, we will see that the Maxwell term contribution is only from a term without $\bar\omega$, 
while the topological term contribution is from the term linear in $\bar \omega$.

Now considering also the metric perturbations, we obtain 
\begin{equation}
\begin{split}
    I \rightarrow \frac{2\alpha L^2}{\kappa_4^2}\int d^3 x  \lim_{z\rightarrow 0} \left[\left( \frac{f^{1/2}-1}{2z^3f^{1/2}} (G_xG_x + G_yG_y) + \frac{q}{2}(A_xG_x+A_yG_y) +\right.\right.\\ \left.\left.
     -\frac{1}{8z^2}
    (G_xG_x^\prime+G_yG_y^\prime) + \frac{f}{2}(A_xA_x^\prime + A_yA_y^\prime)\right)+2i\bar 
    \omega W \epsilon^{ab}A'_a A_b\right]\;,
\end{split}    \label{quadr}
\end{equation}
which, except for the new extra term with $W$ matches the result in  \cite{Hartnoll:2007ai}.

Expanding in frequency Fourier modes for the fluctuations, for example
\begin{equation}
    A_x^0 (t) = \int_{-\infty}^\infty \frac{d\omega}{2\pi}A_x^0(\omega) e^{-i\omega t}\;,
\end{equation}
and considering the on-shell linearized fluctuations (\ref{onshellfluct}), we obtain 
\begin{equation}
\begin{split}
I_{\rm quadratic}=    \frac{i}{2} \int \frac{d\omega}{2\pi} dx^2\omega  
(A_x^0(\omega)A_y^0(-\omega)-A_y^0(\omega)A_x^0(-\omega))\left(\frac{q}{h}-W\right)+\\
    \frac{-3i(1+h^2+q^2)}{4h} \int \frac{d\omega}{2\pi} dx^2\omega  (A_x^0(\omega)G_y^0(-\omega)-A_y^0(\omega)G_x^0(-\omega))+\\
    \frac{9iq(1+h^2+q^2)^2}{32(h^2+q^2)} \int \frac{d\omega}{2\pi} dx^2\omega  (G_x^0(\omega)G_y^0(-\omega)-G_y^0(\omega)G_x^0(-\omega))+\\
    \frac{i(-3+h^2+q^2)^2}{32(h^2+q^2)} \int \frac{d\omega}{2\pi} dx^2\omega  (G_x^0(\omega)G_x^0(-\omega)-G_y^0(\omega)G_y^0(-\omega))\;,
\end{split}    \label{quadract}
\end{equation}
which is now in the form where we can use the recipe for calculating holographic retarded Green's 
functions. Note that the leading terms of order ${\cal O}(\bar\omega^0)$ have cancelled, and the 
only contributions are: from the mixing of terms alluded before ($G_x$ containing $A_y^0$, etc.) for 
the Maxwell and gravity terms, leading to an extra $\bar\omega$, but from the leading term in the 
topological one, and seeing as we already had a $\bar \omega$, this  is of the same order. 

The Green's functions for the current-current, current-energy momentum tensor and two 
energy-momentum tensors are thus 
\bea
 G^R_{J_a,J_b} &= & -i\omega\epsilon_{a,b}\frac{q}{h} =  -i\omega\epsilon_{a,b}
 \left(\frac{\rho}{B}-W\right)\;,\cr
    G^R_{J_a,T_{tb}} &= & -i\omega\epsilon_{a,b}\frac{3(1+h^2+q^2)}{2h} = -i\omega\epsilon_{a,b}\frac{\epsilon}{2B}\;,\cr
G^R_{T_{ta},T_{tb}} &=&  -i\omega\epsilon_{ab}\frac{9q(1+h^2+q^2)^2}{32(h^2+q^2)} +
    i\omega\delta_{ab}\frac{(-3+h^2+q^2)^2}{32(h^2+q^2)}\;,
\eea
where we have written the results in terms of the magnetic field $B=h\a^2$, the charge density 
$\rho=q \a^2$ and the energy density $\epsilon=(1+h^2+q^2)$ of the boundary theory. 

Then finally, using the Kubo formulas, we find the static electric conductivity as 
\begin{equation}
       \sigma_{ab} = -\lim_{\omega \rightarrow 0}\frac{\text{Im}G_{J_aJ_b}^R }{\omega} 
       = \epsilon_{ab}\left(\frac{\rho}{B}-W\right)\;,
\end{equation}
while for the static thermo-electric and heat conductivities we must take into account the magnetization term, 
as already noted, obtaining 
\bea
\alpha_{ab} &=& -\lim_{\omega \rightarrow 0}\frac{\text{Im}G_{T_{ta}J_b}^R}{\omega} + \frac{M}{T}\epsilon_{ab} = \frac{s}{B}\epsilon_{ab},\cr
    \bar{\kappa}_{ab} &=& -\lim_{\omega \rightarrow 0}\frac{\text{Im}G_{T_{ta}T_{tb}}^R}{\omega} + \frac{2(M^E-\mu M)}{T}\epsilon_{ab}.
\eea

\subsection{Anisotropy}

In order to break isotropy, we consider two parameters $(k_x,k_y)$ multiplying the boundary 
space coordinates, giving a metric 
\be
\frac{1}{L^2}ds^2 = -\frac{\alpha^2}{z^2}f(z)dt^2+\frac{\alpha^2}{z^2}(k_xdx^2+k_ydy^2)+\frac{1}{z^2}\frac{dz^2}{f(z)}.
\ee

Note that then we obtain $ \sqrt{-g} = \sqrt{k_xk_y}L^4\alpha^3/z^4$. 

From the equations of motion for gravity, we obtain the condition 
\be
k_x=1/k_y.
\ee

Nothing new is obtained from this moment on, and the analysis proceeds as before, with no new 
physics. We have shown this, however, since in the next section, when we have varying $W(z)$, there
will be new results.

\subsection{Conductivity from entropy function via attractor mechanism and membrane paradigm}

The conductivity can also be obtained using the attractor mechanism, for calculations at the 
horizon, via Sen's entropy function \cite{Sen:2005wa}. The application of Sen's entropy function 
formalism to holography was considered in \cite{Astefanesei:2007vh}  (see also \cite{Astefanesei:2011pz}), 
while the application to the calculation of conductivity was done in \cite{Erdmenger:2016wyp}, 
and was also used in \cite{Alejo:2019utd} for the case of metrics with charge density and 
magnetic field, defined only near the horizon. The fact that we can calculate the conductivity at the horizon, 
as well as at its natural AdS/CFT location, the boundary, is related to the application of the membrane 
paradigm to the AdS black holes, as argued initially by Iqbal and Liu \cite{Iqbal:2008}.

The power of the attractor mechanism is, therefore, that one can both calculate the conductivities only at the 
horizon, and nevertheless obtain them as functions of only the charges of the black hole (defined at infinity), 
related to charges in the field theory. In a top-down (completely known) solution, that is perhaps not too 
impressive, nevertheless we show that this works as expected, since it confirms its use in cases in which the 
solution is only known near the horizon, yet we can still write the conductivities as functions of the charges defined 
at infinity (or in the field theory). 

The near-horizon geometry of an {\em extremal} 4-dimensional 
planar black hole (such as obtained in the attractor 
mechanism) is $AdS_2\times\mathbb{R}^2$, written as 
\be
ds^2 = -vr^2\,dt^2 + w (dx^2+dy^2) + \frac{v}{r^2} dr^2\;,
\ee
where $r=1/z$. Note that this is for a planar black hole, but the formalism works as well as for the 
spherical ($S^2$ instead of $\mathbb{R}^2$) black hole case \cite{Sen:2005wa}. Also, sometimes
$v$ and $w$ are denoted by $v_1$ and $v_2$, in order to emphasize their similarities.

The values of the magnetic and electric fields at the horizon are $F^A_{xy}=B^A$ 
(soemetimes denoted $p^A$) and $F^A_{zt}=e_A$, 
and for the scalars we have the horizon values $\phi_s=u_s$. 

Defining the function $f$ as the integral over the horizon of the Lagrangian density, 
\be
 f(u_s,v,w,B^A,e_A) = \int dxdy \sqrt{-g}\mathcal{L}\;,
\ee
as shown by Sen, the Einstein equations imply that the $u_s,v$ and $w$ are extrema of $f$, while 
the charges $Q_A$ can be defined as its variations with respect to the electric fields, so 
\be
\frac{\partial f}{\partial u_s} =  0\; ,\;\;\;
        \frac{\partial f}{\partial v} =  0\; ,\;\;\;
        \frac{\partial f}{\partial w} =  0\; ,\;\;\;
        \frac{\partial f}{\partial e_A} =  Q^A.
\ee

Then Sen's entropy function is 
\be
\mathcal{E} (u_s,v,w,e_A,B^A;Q^A) = 2\pi (e_A Q^A - f(u_s,v,w,e_A,B^A))\;,
\ee
and is value {\em at its extremum}, defined by the {\em attractor equations}
\be
 \frac{\partial \mathcal{E}}{\partial u_s} =  0\; ,\;\;\;
    \frac{\partial \mathcal{E}}{\partial v} =  0 \;,\;\;\;
    \frac{\partial \mathcal{E}}{\partial w} =  0 \;,\;\;\;
    \frac{\partial \mathcal{E}}{\partial e_A} =  0.
\ee

In our case we don't have a scalar, and we have a single electromagnetic field, with electric field $e$,
so we only have 3 attractor equations. The function $f$ in our case is then (with $4\pi G_N=1$)
\bea
- f &= &  \int dx\,dy\;  \sqrt{-g} \left(-\frac{1}{4}R + \frac{L^2}{4}F_{\mu\nu}F^{\mu\nu} 
- \frac{3}{2}\frac{1}{L^2} + W \frac{L^2}{4}F_{\mu\nu}\tilde{F}^{\mu\nu}\right)\cr
    &= & \frac{1}{2}\int dxdy \left( \frac{B^2L^2v}{w} + w\left(1- \frac{3v}{L^2}
    -\frac{e^2L^2}{v}\right) + 2BeL^2 W \right)\;,
\eea
and the entropy function is 
\be
 \mathcal{E} = 2 \pi  e(\tilde{Q}-BL^2W) + \pi \frac{B^2L^2v}{w} + \pi  w
\left(1- \frac{3v}{L^2}-\frac{e^2L^2}{v}\right)\;,
\ee
where $\tilde{Q} = Q/\text{Vol}$. The attractor equations are then
\bea
     w^2 \left(3-\frac{e^2 L^4 }{v^2}\right)-B^2 L^4 &=&0\cr
    L^2 \left(\frac{B^2 v}{w^2}+\frac{e^2}{v}\right)+\frac{3 v}{L^2}-1&=&0\cr
    \frac{L^2 \left(e w-B v W\right)}{v}-\tilde{Q}&=&0.
\eea

We use the last equation to solve for $e$ in terms of $\tilde Q$ (inverting it), 
and the other two to solve for $v$ and $w$. Then the parameters $v,w, e$ of the ansatz are 
written in terms of the charge density $\tilde Q$ and the magnetic field $B$, as well as the parameters 
$L$ and $W$ in the action, as 
\bea
    v &= & \frac{L^2}{6}\;,\cr
    w &= & \frac{\sqrt{B^2 L^4 \left(W^2+1\right)-2 B L^2 \tilde{Q} W+\tilde{Q}^2}}{\sqrt{3}}\;,\cr
    e &= & \frac{B L^2 W-\tilde{Q}}{2 \sqrt{3} z^2 \sqrt{B^2 L^4 \left(W^2+1\right)-2 B L^2 
    \tilde{Q} W+\tilde{Q}^2}}.
\eea

Since the charge density of the field theory, $\rho$, is dual (couples) to $A_t$, we have
\be
\rho = \frac{\delta S}{\delta A_{t}}\;,
\ee
which gives, on the solution, the same result as the charge density parameter of the black hole,
\be
\rho =  \sqrt{-g}(F^{tz}+W \tilde{F}^{tz})=  \frac{\tilde{Q}}{L^2}.
\ee

For the entropy density, the extremum of the entropy function gives the same as the Hawking formula,
\be
s =   \frac{\sqrt{B^2 L^4 \left(W^2+1\right)-2 B L^2 \tilde{Q} W+\tilde{Q}^2}}{4\sqrt{3}G_N}
=\frac{w}{4G_N}\;,\label{entropyf}
\ee
and this is identified with the entropy density in the field theory.

We can now use these formulas for the entropy density $s$ and the charge density $\rho$ of the 
field theory in the formulas already derived, from Kubo formulas, 
for the electric and thermoelectric conductivities, to find 
these as functions of the charges of the black hole. We find 
\bea
\sigma_{ab} &=&  \epsilon_{ab}\left(\frac{\rho}{B}-W\right) = \epsilon_{ab}
\left(\frac{\tilde{Q}}{BL^2}-W\right)\;,\cr
\alpha_{ab}  &=& \epsilon_{ab}\frac{\sqrt{B^2 L^4 \left(W^2+1\right)-2 B L^2 \tilde{Q} W
+\tilde{Q}^2}}{4\sqrt{3}BG_N}.
\eea

We note that now, with the holographic transport coefficients written in terms of $\tilde Q$, $B$ and $W$
via the attractor mechanism, the explicit $W$ dependence becomes more complicated.

\subsection{S-duality}

As observed in  \cite{Alejo:2019utd}, and made more precise in \cite{Melnikov:2020ktj},
the general formulas for transport coefficients obtained from fluctuations around a solution near the 
horizon of a black hole have an action of S-duality ($Sl(2;\mathbb{Z})$) on 
them, coming from the S-duality invariance of the gravitational action. For a general $Z(\phi)$ (term 
in front of the Maxwell action), the invariance is under 
\bea
F_{\mu\nu}&\rightarrow & Z \tilde F_{\mu\nu}-WF_{\mu\nu}\cr
Z&\rightarrow& -\frac{Z}{Z^2+W^2}\cr
W&\rightarrow & \frac{W}{Z^2+W^2}.\label{SZW}
\eea

This led to the duality relation on the complex conductivity $\sigma\equiv \sigma_{xy}+i\sigma_{xx}$,
\be
\sigma'=-\frac{1}{\sigma}.
\ee

In our case, this is less obvious, since we have effectively fixed $Z$ to 1, but the duality is there.
 
In the case of the dyonic black
hole of  \cite{Hartnoll:2007ai,Hartnoll:2007ih}, which gives a subset of the formulas in  \cite{Alejo:2019utd},
the action of S-duality was hard to understand, as the only relevant limit is the one that takes 
$\rho\rightarrow 0$, followed by $s\rightarrow 0$ 
in the transport coefficients, and previously there was nothing left. 

With the introduction of the $W$ term in this dyonic black hole calculation, and the associated entropy 
function result, it becomes relevant to take the limit $\rho\rightarrow 0$, and obtain 
\be
\sigma_{xx}=0\;,\;\;\; \sigma_{xy}=-W\;,\;\; \a_{xx}=0\;,\;\;\;\a_{xy}=\frac{s}{B}.
\ee

S-duality then acts nontrivially, as
\be
W\rightarrow \frac{1}{W}\Rightarrow \sigma_{xy}\rightarrow -\frac{1}{\sigma_{xy}}.
\ee

The limit $\rho\rightarrow 0$ is understood, from the point of view of the dyonic black hole, as 
the limit when the electric charge goes to zero, keeping the magnetic charge finite. Considering also the
entropy function calculation leading to (\ref{entropyf}), we obtain that in this limit, 
\be
\a_{xy}=\frac{s}{B}=\frac{c}{\sqrt{3}}\sqrt{W^2+1}\;,\label{alphaxy}
\ee
where 
\be
c=\frac{L^2}{4G_N}=\frac{\pi Z}{g_4^2}
\ee
is the central charge of the dual field theory, with the second form being due to our fixing $Z/g_4^2$.

But, moreover, if we would keep $Z$ free, we would obtain $\sqrt{W^2+Z^2}$ in (\ref{alphaxy}), 
which is an S-duality invariant, see (\ref{SZW}).

\section{Radially varying topological term $W(z)$}

In this section we consider a more general model, with a field dependent topological term, so an 
Einstein-Maxwell-dilaton model with a nontrivial dilaton, as considered for instance in 
\cite{Alejo:2019utd,Melnikov:2020ktj},
\begin{equation}
\begin{split}
        I = \frac{2}{\kappa_4^2}\int d^4 x \sqrt{-g} \left( -\frac{1}{4}R \right. & 
        -\frac{1}{2}\left[(\partial\phi )^2 
        +\Phi(\phi)((\partial \chi_1)^2 + (\partial \chi_2)^2)\right]-V(\phi ) +  \\ 
        &   \left. + Z(\phi )\frac{L^2}{4}F_{\mu\nu}F^{\mu\nu}   + W(\phi ) \frac{L^2}{4}F_{\mu\nu}\tilde{F}^{\mu\nu} \right) .
\end{split}        \label{genact}
\end{equation}

The reason for introducing $W(\phi)$ is to have a nontrivial contribution to the Einstein equation
(and the Maxwell equation); otherwise, as we saw in the previous section, this is a topological term. 

Here the axions $\chi_i$ are introduced to have a breaking of translational invariance of the theory, 
via a linear axion ansatz
\bea
\chi_1&=& k_1 x\cr
\chi_2&=& k_2 y.
\eea

We will go back to the isotropic case by considering $k_1=k_2=k$. 

In order for the ansatz above to be consistent with the axion equations of motion, 
\be
 \Phi(\phi) \partial_\mu\partial^\mu \chi_{(1,2)} + \Phi^\prime(\phi)\partial_\mu\chi_{(1,2)}
 \partial^\mu\phi=0\;,
\ee
we will assume that the dilaton is static and only depends on the radial direction, so $\phi=\phi(z)$. 

The equation of motion for the dilaton is 
\be
\partial_\mu\partial^\mu\phi - V^\prime (\phi) - \frac{1}{2}\Phi^\prime(\phi)\left[(\partial \chi_1)^2 
+ (\partial \chi_2)^2\right] + \frac{L^2}{4}(Z^\prime(\phi)F_{\mu\nu} + W^\prime(\phi)
F_{\mu\nu}\tilde{F}^{\mu\nu})=0. 
\ee

To guarantee that we have a solution, we must impose that 
\bea
V(0) & =& \frac{-6}{L^2}\;,\cr
V^\prime(0) & =& 0.
\eea

\subsection{Set-up}

However, the ansatz considered so far is still too complicated to solve, so instead of the arbitrary 
functions of the dilaton $W(\phi), \Phi(\phi), V(\phi)$ and $Z(\phi)$, on top of the radially varying dilaton 
$\phi(z)$, we will simplify further and directly consider
independent functions of the radial coordinate $z$, so $W(z), \Phi(z), V(z)$ and $Z(z)$. 
At this time, we could not find a way to solve the more interesting case of arbitrary (given functions) $W(\phi), \Phi(\phi), 
V(\phi), Z(\phi)$ and solving for $\phi(z), F_{\mu\nu}$ and $g_{\mu\nu}$, so we are restricting to this simple case. 
The resulting 
reduced Einstein-Maxwell-dilaton model is 
\begin{equation}
\begin{split}
        I = \frac{2}{\kappa_4^2}\int d^4 x \sqrt{-g} \left( -\frac{1}{4}R \right. & 
        -\frac{1}{2}\left[(\partial\phi )^2 
        +\Phi(z)((\partial \chi_1)^2 + (\partial \chi_2)^2)\right]-V(z) +  \\ 
        &   \left. + Z(z )\frac{L^2}{4}F_{\mu\nu}F^{\mu\nu}   + W(z ) \frac{L^2}{4}F_{\mu\nu}\tilde{F}^{\mu\nu} \right) .
\end{split}        
\end{equation}

Now the gauge field equation of motion becomes
\be
\frac{1}{\sqrt{-g}}\partial_\mu \sqrt{-g}\left(Z(z )F^{\mu\nu} + W(z )\tilde{F}^{\mu\nu} \right) = 0.
\ee

We take the same ansatz for the metric as in the previous section, 
\be
 \frac{1}{L^2}ds^2 = -\frac{\alpha^2}{z^2}f(z)dt^2
 +\frac{\alpha^2}{z^2}(dx^2+dy^2)+\frac{1}{z^2}\frac{dz^2}{f(z)}\;,
\ee
where $f(z)$ is the same function, 
\be
f(z) = 1+(h^2+q^2)z^4-(1+h^2+q^2)z^3\;,
\ee
and the ansatz for the field strength is also unchanged, 
\be
F=h\a^2dx\wedge dy +q\a dz\wedge dt\Rightarrow 
F_{\mu\nu} = 
\begin{pmatrix}
0 & 0 & 0 & -q\alpha\\
0 & 0 & h\alpha^2 & 0\\
0 & -h\alpha^2 & 0 & 0\\
q\alpha & 0 & 0 & 0
\end{pmatrix}.
\ee

Then the Maxwell equation of motion gives
\be
h W^\prime (z) - q Z^\prime (z) = 0\;,\label{Mxeq}
\ee
solved by 
\be
Z(z)=\frac{h W(z)}{q}+Z_0\;,\label{Zsol}
\ee
where $Z_0$ is a constant. 

The Einstein equations,
\be
 R_{\mu\nu}  =  K_{\mu\nu} + \frac{1}{2}g_{\mu\nu}V(z ) + Z(z ) \left( 2L^2F_{\mu\sigma}
 F_\nu^\sigma - \frac{L^2}{2}g_{\mu\nu}F_{\sigma\rho}F^{\sigma\rho} \right) \;,
\ee
give 3 independent equations, which can be taken to be the $xx, zz$ and $xy$ components, for instance, 
and can be used to fix $K_{zz}(z)$, $K_{xx}(z)=K_{yy}(z)$ and $V(z)$\;,
where 
\be
K_{zz} = \frac{1}{2}(\partial_z\phi )^2\;,\;\;\;
K_{xx} = \frac{1}{2}\Phi(z) k_1^2\;,\;\;\;
K_{yy} = \frac{1}{2}\Phi(z) k_2^2.
\ee

The solution of the equations of motion is then 
\bea
K_{zz}(z) &= & 0\;,\cr
K_{xx}(z) &= & -\frac{2 \alpha ^2 z^2 \left(h^2+q^2\right) (h W(z)+q (Z_0-1))}{q}\;,\cr
V(z) &= & \frac{2 h z^4 \left(h^2+q^2\right) W(z)+2 q \left(z^4 (Z_0-1) \left(h^2
+q^2\right)-3\right)}{L^2 q}.
\eea

Note that, strictly speaking, the above solution means that $\phi$ is constant, so $\Phi, V, Z, W$ should 
have been constant as well. Except, of course, in a correct solution we should have varied $\Phi(\phi), 
V(\phi), Z(\phi), W(\phi)$ with the chain rule to obtain the correct dilaton equation of motion, which 
was not done here. 

So the above must be thought of as a simple toy model for the correct case. We have fixed $Z(z), 
\Phi(z), V(z)$ in terms of the independent $W(z)$ from the equations of motion, considered as the 
only variable, set by hand.

\subsection{Fluctuations}

To calculate the quadratic action for fluctuations around the background solution, and find the 
holographic Green's functions and the transport coefficients through Kubo formulas, we proceed as in the 
previous section. 

We add fluctuations to the off-diagonal metric and gauge field in the spatial boundary directions, 
\bea
g_{tx} & =& \frac{\alpha  \epsilon  G_x(z) e^{-i t \omega }}{z^2}\;,\cr
    g_{ty} & =& \frac{\alpha  \epsilon  G_y(z) e^{-i t \omega }}{z^2}\;,\cr
    dA_{x} & =& \epsilon  A_x(z) e^{-i t \omega }\;,\cr
    dA_{y} & =& \epsilon  A_y(z) e^{-i t \omega }\;,
\eea
and consider the same ansatz with infalling boundary conditions at the horizon, 
\bea
A_x(z) & =& f(z)^\nu a_x(z)\;,\cr
G_x(z) & =& f(z)^{(1+\nu)} g_x(z).\label{incans}
\eea

At linear level, we now obtain the equations of motion 
\bea
i \bar{\omega} G_x'(z) &= & -\frac{4 z^2 (h W(z)+q Z_0)}{q} \left[-i q \bar{\omega} A_x(z)\right.\cr
&&\left.+h (z-1)A_y'(z) \left\{z \left(z \left[z \left(h^2+q^2\right)-1\right]-1\right)-1\right\}
+h q G_y(z)\right]\;,\cr
&& q (z-1) \left\{z \left(z \left[z \left(h^2+q^2\right)-1\right]-1\right)-1\right\} \times\cr
&&\times\left[4 z^3 A_x'(z) (h W(z)+q
Z_0)-2 G_x'(z)+z G_x''(z)\right]\cr
& =& 4 i h \bar{\omega} z^3 A_y(z) (h W(z)+q Z_0)+4 q z^3
G_x(z) \left(h^2-h q W(z)-q^2 (Z_0-1)\right).\cr
&&
\eea

We expand again the solutions in orders of     $\bar{\omega}$:
\bea
a_x(z) &=&  a_x^0(z) + \bar{\omega} a_x^1(z)+...\cr
g_x(z) &=&  g_x^0(z) + \bar{\omega} g_x^1(z)+...
\eea

We solve the equations of motion order by order in $\epsilon$ and in $\bar \omega$, as before, focusing 
on the linearized perturbations (first order in $\epsilon$).
At zeroth order in $\bar\omega$, we obtain 
\bea
a_x^{0\prime} (z)   &= & -q g_x^0(z)\;,\cr
g_x^{0\prime\prime} (z) &= & -\frac{2 g_x^{0\prime}(z) \left(3 h^2 z^4-2 h^2 z^3+3 q^2 z^4
-2 q^2 z^3-2 z^3-1\right)}{(z-1) z \left(h^2 z^3+q^2z^3-z^2-z-1\right)}\cr
   & =& -2\frac{\psi^\prime(z)}{\psi(z)}g_x^{0\prime}(z)\;,
\eea
which matches what we got in the previous section for constant $W$, as expected. So the same solution 
as before is also valid now, 
\be
a_x^0(z) = \alpha_x -qz\gamma_x
    g_x^0(z) = \gamma_x\;,
\ee
where $\gamma_x$ and $\a_x$ are constants. 

At first order in $\bar\omega$, we obtain the equation for $a_x$
\be
 a_x^{1\prime} (z) = \frac{\theta(z)+\gamma (z) + \psi (z)}{4 h (z-1) z^2 \left(z \left(z \left(z
   \left(h^2+q^2\right)-1\right)-1\right)-1\right) (h W(z)+q Z_0)}\;,
\ee
where we have defined the functions
\bea
\theta(z) &=& -4 q h (z-1) z^2 g_x^1(z) \left(z \left(z \left(z \left(h^2+q^2\right)-1\right)-1\right)-1\right) (h
W(z)+q Z_0)\;,\cr
   \gamma(z) &=& q\frac{4 i z^2 (h W(z)+q Z_0) }{q \left(h^2+q^2-3\right)} 
   \left(h^3 z^2 (4 z-3) (\gamma_x q z-\alpha_x)+h^2
   q (\gamma_y q z-\alpha_y) + \right.\cr
& & \left.+h z^2 \left(q^2 (4 z-3)-3\right) (\gamma_x q z-\alpha_x)+q   \left(q^2-3\right) 
(\gamma_y q z-\alpha_y)\right)\;,\cr
   \psi (z) &=& -qi \gamma_y z^2
   \left(4 z \left(h^2+q^2\right)-3 \left(h^2+q^2+1\right)\right).
\eea

There is also a similarly complicated equation for $g_x^1$. But it turns out that we can write both these 
equations (for $a_x$ and $g_x^1$) in the same form as we did in the previous section, 
\bea
a_x^{1\prime} (z) + q g_x^1(z) &= & \mathcal{A}^0_x(z)\;,\cr
g_x^{1\prime\prime} (z) +2\frac{\psi^\prime(z)}{\psi(z)}g_x^{1\prime}(z)  &= & \mathcal{G}^0_x (z)\;,
\eea
except with different $ \mathcal{G}^0_x$ and $\mathcal{A}^0_x(z)$. Solving the equations in the 
same way (see \cite{Hartnoll:2007ai}), we obtain 
\be
g_x^{1\prime} (z) = \frac{c_2}{\psi (z)^2}+ \frac{1}{\psi (z)^2} \int_0^z \mathcal{G}^0_x (u)
\psi (u)^2 du.
\ee

This again blows up at $z=1$, so we must impose regularity by putting $c_2=0$, and 
\be
\int_0^1 \mathcal{G}^0_x (u)\psi (u)^2 du = 0\;,
\ee
which relates $\gamma_x$ to the other integration constants, 
\begin{equation}
\begin{split}
    \gamma_x = &  -\frac{\left(h^2+q^2-3\right) \left(\gamma_y q \left(2 Z_0 \left(h^2+q^2\right)+h^2+q^2+3\right)-4
   \alpha_y Z_0 \left(h^2+q^2\right)\right)}{3 h \left(h^2+q^2+1\right)}+\\
   & \quad\quad\quad + \frac{i \left(h^2+q^2-3\right)}{3 \left(h^2+q^2+1\right)} \int_0^1 -\frac{4 i \left(h^2+q^2\right) W(z) (\gamma_y q z-\alpha_y)}{q}
   \, dz\;,
\end{split}
\end{equation}
and similarly for $\gamma_y$. Therefore we have the solution
\bea
    a_x^{1} (z) & =& \alpha_x -  q \int_0^z g_x^1(u)du -i\int_0^z \mathcal{A}^0_x(u)du\;,\cr
    g_x^{1} (z) & =& \gamma_x - i\int_0^z  \frac{1}{\psi (u)^2} \gamma_x\mathcal{W}(u) du\;,
\eea
where $\mathcal{W}(u)$ is a function depending on the parameters of the solution, which we write,
together with $\mathcal{A}^0_x(u)$), in Appendix A.

Next one would need to calculate the quadratic action as we did in (\ref{quadr}) and (\ref{quadract}), and 
extract the transport coefficients, but it is now too involved (one could do numerics for it, but we 
leave that for further work). 

Note that, as before, besides the solutions with ansatz (\ref{incans}), with $\pm \nu$, we also have 
a constant solution, 
\bea
A_x & =& \delta_x\;,\;\;\;
    G_x  = -\frac{i \bar{\omega}}{h}\delta_x\;,\cr
    A_y & =& \delta_y\;,\;\;\;
    G_y  = \frac{i \bar{\omega}}{h}\delta_y.
\eea

The solutions become near the boundary at $z=0$
\bea
A_x^0 & =& \delta_x + \alpha_x(\gamma_x,\gamma_y)\;,\cr
G_x^0 & =& -\frac{i \bar{\omega}}{h}\delta_x + \gamma_x\;,\cr
A_y^0 & =& \delta_y + \alpha_y(\gamma_x,\gamma_y)\;,\cr
G_y^0 & =& \frac{i \bar{\omega}}{h}\delta_y+ \gamma_y.
\eea

\subsection{Anisotropy}

We can introduce anisotropy as in the previous section, via a metric ansatz with $k_x\neq k_y$, 
\be
\frac{1}{L^2}ds^2 = -\frac{\alpha^2}{z^2}f(z)dt^2+\frac{\alpha^2}{z^2}(k_xdx^2+k_ydy^2)+\frac{1}{z^2}\frac{dz^2}{f(z)}\;,
\ee
and now it is not trivial anymore. The Einstein equations give now different values for $K_{xx}$ 
and $K_{yy}$, and we find 
\bea
 V(z) & =& \frac{2 \left(h z^4 W(z) \left(h^2+k_x k_y q^2\right)+q \sqrt{k_x k_y} 
 \left(h^2 z^4 Z_0-k_x k_y \left(z^4 \left(h^2-q^2
   Z_0+q^2\right)+3\right)\right)\right)}{L^2 q (k_x k_y)^{3/2}} \;,\cr
K_{xx}(z) & =& -\frac{2 \alpha ^2 k_x z^2 \left(W(z) \left(h^3+h k_x k_y q^2\right)
+q \sqrt{k_x k_y} \left(h^2 (Z_0-k_x
   k_y)+k_x k_y q^2 (Z_0-1)\right)\right)}{q (k_x k_y)^{3/2}}\;,\cr
   K_{yy} & =& -\frac{2 \alpha ^2 k_y z^2 \left(W(z) \left(h^3+h k_x k_y q^2\right)+q 
   \sqrt{k_x k_y} \left(h^2 (Z_0-k_x
   k_y)+k_x k_y q^2 (Z_0-1)\right)\right)}{q (k_x k_y)^{3/2}}.\cr
   &&
\eea

Repeating the procedure from the previous subsection, we obtain the $\gamma_x$ and 
$\gamma_y$ given in Appendix B.

\subsection{Conductivity from entropy function}

Considering the same $AdS_2\times \mathbb{R}^2$ ansatz for the near-horizon metric of the 
planar extremal black hole in 4 dimensions, 
\be
ds^2 = -\frac{v}{z^2}(dt^2 - dz^2) + w(dx^2 + dy^2)\;,
\ee
we compute as before the boundary spatial integral of the Lagrangian density, 
\be
 f =  \int dxdy \int d^4 x \sqrt{-g}\mathcal{L}\;,
\ee
and finally Sen's entropy function, which becomes
\be
\mathcal{E} = \frac{\pi }{\alpha ^2 v w z^2} \Gamma + \frac{\pi}{\alpha ^2 e_A v w^2 z^2}\Theta\;,
\ee
where 
\begin{equation}
\begin{split}
    \Gamma =  & 4 \alpha ^4 h^2 v w z^2 \left(w^2-L^2 v Z_0\right)-4 e^2 w^2 z^4 
    \left(v-L^2 Z_0\right) + \\
   & + \alpha ^2 (L^2 Z_0 \left(-B^2 v^2+e^2 w^2 z^2 \left(z^2-4 w\right)+4 h^2 v^2 z^4\right)+\\
   & +v w \left(2 e z^2
   \left(2 e w^2+Z_0\right)-4 h^2 w z^4-13 w\right))
\end{split}   
\end{equation}
and
\begin{equation}
\begin{split}
    \Theta = & \alpha  L^2 v W(z) (e^2 w^2 z^2 \left(-2 \alpha  B+\alpha ^2 h \left(z^2-4 w\right)
    +4 h z^2\right)+\\
    & -\alpha ^2 h v^2
   \left(B^2-4 h^2 z^2 \left(z^2-\alpha ^2 w\right)\right)).
\end{split}   
\end{equation}

The attractor equations are, as before
\be
\frac{\partial \mathcal{E}}{\partial v} =  0 \;,\;\;\;
    \frac{\partial \mathcal{E}}{\partial w} =  0 \;,\;\;\;
    \frac{\partial \mathcal{E}}{\partial e_A} =  0.
\ee

The full solutions for $v, w$ and $e$ are too big to be shown here, though it should be possible to 
find them numerically.

\subsection{S-duality}

One important reason to consider the more general action (\ref{genact}) is S-duality. As explained in the 
previous section, such an action is manifestly invariant under S-duality acting on $Z(\phi) $ and 
$W(\phi)$. 

However, we considered the ``toy model'' with only $z$ dependence for $W(z), Z(z), V(z)$, $\Phi(z)$, so 
we need to check S-duality on the solutions. 
Of course, we see that the Maxwell equation (\ref{Mxeq}) is indeed S-duality invariant, 
but in order to have the solutions be as well, we need that $Z_0=0$ in (\ref{Zsol}).

Since $W(z)$ and $Z(z)$ are functions of the radial coordinate $z$, now we have to ask: at what position 
$z$ is the action of S-duality relevant to transport coefficients to be considered? 
On the one hand, by virtue of calculating the holographic Green's functions and using the Kubo formulas
at the boundary, that is where it seems we should consider them. But on the other hand, the conductivity
calculated from Sen's entropy function in the attractor mechanism is obtained at the horizon, so that is 
where it seems to be needed in this case. 

The two calculations are related by the application to AdS/CFT of the membrane paradigm, as done 
by Iqbal and Liu, and as shown for instance in \cite{Alejo:2019utd}, but in the case of our toy model, 
that is guaranteed by the fact that $Z(z)$, as well as $V(z)$ and $\Phi(z)$, are related to $W(z)$, and 
$Z(z)$ is related to it via a duality-invariant proportionality relation, 
\be
Z(z)=\frac{h}{q}W(z)\;,
\ee
which is one reason why the toy model set-up is a sensible one.

\section{Conclusions}

In this paper we have considered the introduction of a topological term $W F_{\mu\nu}\tilde F^{\mu\nu}$
in the 3+1 dimensional 
gravitational action for the Einstein-Maxwell model used to holographically calculate transport 
coefficients in strongly coupled 2+1 dimensional materials, using a dyonic black hole background that 
asymptotes to AdS space. 

Considering first a constant $W$, 
we have found that the results match the general results  in \cite{Alejo:2019utd}
from the calculation in an a priori unknown 
metric, at the horizon of the black hole. Using also the attractor mechanism and Sen's entropy function, 
the transport coefficients were written in terms of $\rho=\tilde Q$, $B$ and $W$, and we have found 
the action of S-duality on them. 

A toy model for the complicated case of a general solution with nontrivial $W(\phi), Z(\phi)$, $V(\phi), 
\Phi(\phi)$ was the case with functions of the radial coordinate $z$, $W(z), Z(z), V(z)$ and $\Phi(z)$. 
We obtained the solutions for fluctuations, which were very complicated, so we did not proceed to 
find the transport coefficients, though those could be found numerically from our results. 
We have also shown how to introduce anisotropy in this case, and found the solutions for fluctuations 
in this case. We have set up the case of the attractor mechanism and Sen's entropy function, 
again leaving the complete (and very ugly) formulas for later. S-duality arguments sharpened the 
idea of the toy model, as well as its relevance. 

There are many things left for further work, for instance the numerical evaluation of the transport 
coefficients in the toy model with $z$ dependence. This is also viewed as a first step towards a 
case based on a more complete solution, involving not only a dyonic black hole, but nontrivial $\phi(z)$
and $W(\phi), Z(\phi), V(\phi), \Phi(\phi)$.

\section*{Acknowledgements}

We thank Dmitry Melnikov for useful discussions. The work of HN is supported in part by  CNPq grant 301491/2019-4 and FAPESP grant 2019/21281-4. HN would also like to thank the ICTP-SAIFR for their 
support through FAPESP grant 2016/01343-7.
The work of CLT is supported by CNPq grant 	141016/2019-1.

\appendix

\section{Solutions for fluctuations in the case of $W(z)$}

\begin{equation}
\begin{split}
   \mathcal{W}(u) = & -\frac{i}{\gamma_x}  \int_0^u -i \left(\frac{4 \left(h^2+q^2\right) W(z) (\gamma_y q z-\alpha_y)}{q} 
        \right) \, dz \\
   & -u \frac{\gamma_x h^3 \left(u^2 \left(2 q^2 (u-1) (4 u-3)-7 u+6\right)+4\right)}{\gamma_x h
   \left(h^2+q^2-3\right)} \\
   & -u \frac{2 h^2 \left(\gamma_y q^3 (3-2 u)+\left(2 q^2-3\right)
   Z_0 (\gamma_y q u-2 \alpha_y)+3 \gamma_y q (u-1)\right)}{\gamma_x h
   \left(h^2+q^2-3\right)} \\
   & -u \frac{h^4 (\gamma_y q (2 u (Z_0-1)+3)-4 \alpha_y
   Z_0)+\gamma_x h^5 (u-1) u^2 (4 u-3)}{\gamma_x h
   \left(h^2+q^2-3\right)} \\   
   & -u \frac{\gamma_x h \left(q^2 \left(u^2 \left(q^2 (u-1) (4 u-3)-7 u+6\right)+4\right)+3 u^2\right)}{\gamma_x h
   \left(h^2+q^2-3\right)} \\      
   & -u \frac{q
   \left(q^2-3\right) \left(3 \gamma_y+\gamma_y q^2 (2 u (Z_0-1)+3)-4 \alpha_y q Z_0\right)}{\gamma_x h
   \left(h^2+q^2-3\right)}.        
\end{split}        
\end{equation}

\begin{equation}
\begin{split}
    \mathcal{A}^0_x(u) & = 
    i\frac{ (h^3 z^2 (4 z-3) (\gamma_x q z-\alpha_x)+h^2 q (\gamma_y q z-\alpha_y)}{h (z-1) \left(h^2+q^2-3\right) \left(z \left(z \left(z
   \left(h^2+q^2\right)-1\right)-1\right)-1\right)}\\
   & + i\frac{hz^2 \left(q^2 (4 z-3)-3\right) (\gamma_x q z-\alpha_x)+q \left(q^2-3\right) (\gamma_y q
   z-\alpha_y))}{h (z-1) \left(h^2+q^2-3\right) \left(z \left(z \left(z
   \left(h^2+q^2\right)-1\right)-1\right)-1\right)}\\
    & + \frac{i \gamma_y q \left(h^2 (3-4 z)+q^2 (3-4 z)+3\right)}{4 h (z-1) \left(z \left(z \left(z
   \left(h^2+q^2\right)-1\right)-1\right)-1\right) (h W(z)+q Z_0)}
\end{split}
\end{equation}
and similarly for $\mathcal{A}^0_y(u)$.

\section{Solution for fluctuations for the anisotropic model}

\begin{equation}
\begin{split}
    \gamma_y  = \frac{\left(h^2+q^2-3\right)}{3 h k_x \left(h^2+q^2+1\right)} & \left(-i h k_x \int_0^1 \frac{4 i k_y W(z) \left(h^2+k_x k_y q^2\right) (\gamma_x q z-\alpha_x)}{q (k_x k_y)^{3/2}} \, dz  \right.\\
    & \quad \quad\left. -4 \alpha_x Z_0 \left(h^2+k_x k_y q^2\right) \right.\\
    \quad&\left.+\gamma_x q \left(k_x k_y
   \left(h^2+2 q^2 Z_0+q^2+3\right)+2 h^2 Z_0\right)\right)
\end{split}   
\end{equation}
and
\begin{equation}
\begin{split}
    \gamma_x  = \frac{\left(h^2+q^2-3\right)}{3 h k_y \left(h^2+q^2+1\right)}  &\left(i h k_y \int_0^1 \frac{4 i k_x W(z) \left(h^2+k_x k_y q^2\right) (\gamma_y q z-\alpha_y)}{q (k_x k_y)^{3/2}} \, dz\right.\\
    &\quad\quad\left. -4 \alpha_y Z_0 \left(h^2+k_x k_y q^2\right)\right.\\
   & \left.\quad+\gamma_y q \left(k_x k_y
   \left(h^2+2 q^2 Z_0+q^2+3\right)+2 h^2 Z_0\right)\right)
\end{split}   
\end{equation}

\bibliography{dualityCaio}
\bibliographystyle{utphys}

\end{document}